%% file: paper.tex
\documentclass[
12pt,
pra,
a4paper,
amsmath,
superscriptaddress,
showkeys,
floatfix,
notitlepage,
longbibliography
]{revtex4-2}

\usepackage[utf8]{inputenc}
\usepackage[T1]{fontenc}

\input{defines}

\begin{document}

\title{\papertitle}

\author{Piotr Nowakowski}
\thanks{These authors contributed equally to this work.}
\affiliation{Group of Computational Life Sciences, Division of Physical Chemistry, Ru\dj er Bo\v{s}kovi\'c Institute, 10000 Zagreb, Croatia}
\affiliation{Max Planck Institute for Intelligent Systems, Heisenbergstra{\ss}e~3, D-70569 Stuttgart, Germany}
\affiliation{IV. Institute for Theoretical Physics, University of Stuttgart,  Pfaffenwaldring 57, D-70569 Stuttgart, Germany}

\author{Nima Farahmand Bafi}
\thanks{These authors contributed equally to this work.}
\affiliation{Max Planck Institute for Intelligent Systems, Heisenbergstra{\ss}e~3, D-70569 Stuttgart, Germany}
\affiliation{IV. Institute for Theoretical Physics, University of Stuttgart,  Pfaffenwaldring 57, D-70569 Stuttgart, Germany}
\affiliation{Institute of Physical Chemistry, Polish Academy of Sciences, Kasprzaka 44/52, Warsaw, Poland}

\author{Giovanni Volpe}
\affiliation{Department of Physics, University of Gothenburg, SE-41296, Gothenburg, Sweden}

\author{Svyatoslav Kondrat}
\affiliation{Max Planck Institute for Intelligent Systems, Heisenbergstra{\ss}e~3, D-70569 Stuttgart, Germany}
\affiliation{IV. Institute for Theoretical Physics, University of Stuttgart,  Pfaffenwaldring 57, D-70569 Stuttgart, Germany}
\affiliation{Institute of Physical Chemistry, Polish Academy of Sciences, Kasprzaka 44/52, Warsaw, Poland}
\affiliation{Institute for Computational Physics, University of Stuttgart, Allmandring 3, 70569 Stuttgart, Germany}

\author{S. Dietrich}
\affiliation{Max Planck Institute for Intelligent Systems, Heisenbergstra{\ss}e~3, D-70569 Stuttgart, Germany}
\affiliation{IV. Institute for Theoretical Physics, University of Stuttgart,  Pfaffenwaldring 57, D-70569 Stuttgart, Germany}

\begin{abstract}

Critical Casimir forces emerge among particles or surfaces immersed in a near-critical fluid, with the sign of the force determined by surface properties and with its strength tunable by minute temperature changes.
Here, we show how such forces can be used to trap a colloidal particle and levitate it above a substrate with a bull's-eye pattern consisting of a ring with surface properties opposite to the rest of the substrate.
Using the Derjaguin approximation and mean-field calculations, we find a rich behavior of spherical colloids at such a patterned surface, including
sedimentation towards the ring and levitation above the ring (ring levitation) or above the bull's-eye's center (point levitation).
Within the Derjaguin approximation, we calculate a levitation diagram for point levitation showing the depth of the trapping potential and the height at which the colloid levitates, both depending on the pattern properties, the colloid size, and the solution temperature.
Our calculations reveal that the parameter space associated with point levitation shrinks if the system is driven away from a critical point, while, surprisingly, the trapping force becomes stronger.
We discuss the application of critical Casimir levitation for sorting colloids by size and for determining the thermodynamic distance to criticality. 
Our results show that critical Casimir forces provide rich opportunities for controlling the behavior of colloidal particles at patterned surfaces.

\end{abstract}

\date{\today}

\maketitle

\section{Introduction}

Critical Casimir forces act among objects immersed in near-critical fluids, such as binary liquid mixtures near their consolute point \cite{maciolek_collective_2018, dantchev_critical_2023, gambassi_critical_2024, dantchev_casimir_2024}.
Analogous to quantum Casimir forces \cite{gambassi_casimir_2009}, which are due to quantum fluctuations \cite{casimir_attraction_1948, kardar_friction_1999}, these forces emerge due to fluctuations of an order parameter characterizing a critical point. 
The strength of critical Casimir forces depends on the thermodynamic proximity of the fluid to its criticality.
Typically, two surfaces with similar properties, \eg, preferring the same component of a binary liquid mixture, attract each other, while surfaces with opposite properties repel each other \cite{maciolek_collective_2018}.

Critical Casimir forces were predicted by \citeauthor{Fisher1978} \cite{Fisher1978} in \citeyear{Fisher1978}, but directly observed only in \citeyear{hertlein_direct_2008} by \citeauthor{hertlein_direct_2008} \cite{hertlein_direct_2008}.
Since then, they have attracted strong interest from the perspective of basic research \cite{Gambassi2009, Kondrat2009, Mohry2010, paladugu_nonadditivity_2016, Vasilyev2018, FarahmandBafi2020} and as a means of finely controlling interparticle interactions by minute changes of temperature or of other thermodynamic parameters.
For instance, critical Casimir forces have been used to manipulate colloidal phase transitions \cite{nguyen_controlling_2013} as well as structural properties of colloidal suspensions \cite{nguyen_switching_2017} and of deposited nanoparticles \cite{marino_controlled_2021, vasilyev_debye_2021}. One aims at controlling the localization, the orientation, and the movement of microparticles on patterned substrates \cite{wang_nanoalignment_2024-1}, as well as to counteract attractive Casimir--Lifshitz forces \cite{schmidt_tunable_2023}, which can be significant in micro-electro-mechanical systems (MEMS) \cite{chan_quantum_2001}.

The possibility to tune antagonistic boundary conditions at the surface of colloidal particles and at a substrate allows one to realize levitation of particles in terms of critical Casimir forces.
\label{ref:1:Earnshaw}Particle levitation has garnered significant attention \cite{munday_measured_2009, bimonte_reversing_2015, silvera_batista_controlled_2017, bukosky_extreme_2019, carrasco-fadanelli_sedimentation_2023, raynal_shortcuts_2023}, especially given the constraints imposed by Earnshaw's theorem on systems capable of providing particle levitation. Earnshaw's theorem asserts that a system of particles cannot be stabilized using electrostatic forces alone \cite{earnshaw_nature_1848}. This theorem has been extended to encompass systems with arbitrary power-law interactions \cite{jones_earnshaws_1980}, both mobile and immobile charges \cite{rahi_constraints_2010}, and certain cases involving quantum-electrodynamic Casimir interactions \cite{rahi_constraints_2010}.
Recently, particle levitation has been demonstrated experimentally using a direct-current electric field \cite{silvera_batista_controlled_2017}, an oscillatory electric field \cite{bukosky_extreme_2019}, and catalytically active colloids \cite{carrasco-fadanelli_sedimentation_2023}.
However, such methods either require the application of an external field, or they cannot conveniently switch on and off particle levitation.
An exciting application of critical Casimir forces is the possibility to control trapping and levitation of colloids and nanoparticles by minute temperature changes.
Such a possibility has been demonstrated by \citeauthor{Troendle2009} \cite{Troendle2009, Troendle2010, Troendle2011}, who showed theoretically, by using mean field theory (MFT) \cite{Troendle2009,Troendle2010} and the Derjaguin approximation \cite{Troendle2011}, as well as experimentally for a water--lutidine critical mixture \cite{Troendle2011}, that a colloid could levitate above alternating stripes of opposite surface preferences (\ie, stripes preferring water or lutidine). 
While the colloids levitated above the surface at a fixed normal distance from it, they could still move \emph{freely} along the stripes in the lateral direction.

In the present study, we use MFT and the Derjaguin approximation in order to demonstrate fully \emph{localized} (\ie, \emph{point}-like) levitation of a colloidal particle above a substrate endowed with a bull's-eye pattern. 
It is interesting to note that such patterns are used in so-called circular Bragg reflector cavities \cite{scheuer_lasing_2005, sapienza_nanoscale_2015}, for example to produce polarized and un-polarized light emission from a single emitter \cite{peniakov_polarized_2023}.
The bull's-eye pattern consists of a ring with the same surface preference as the colloid, whereas the rest of the substrate has the opposite surface preference (\cref{fig:model}). 
We focus exclusively on critical Casimir interactions and do not consider other forces acting between the colloid and the substrate, such as electrostatic or van der Waals interactions. 
\label{ref:2:other_forces}In the vicinity of a critical point, these forces can be tuned to be weak compared to critical Casimir forces \cite{hertlein_direct_2008}. We thus expect them to shift the levitation position of the colloid only slightly.

Our article is organized as follows. In \cref{sec:model} we discuss the details of our set-up, and in \cref{sec:methods} the methods we use in order to study such systems.
After presenting general considerations about possible configurations of a colloid above the bull's-eye pattern (\cref{sec:res:general}), we compute a configuration diagram by using the Derjaguin approximation (\cref{sec:res:conf_diagram}), and demonstrate point levitation via mean-field calculations (\cref{sec:res:mft}).
In \cref{sec:res:lev_diagram} we use the Derjaguin approximation in order to calculate a levitation diagram, showing the trapping potential and the levitation height of the colloid.
We conclude in \cref{sec:concl}.

\section{Model}
\label{sec:model}

\begin{figure}
\begin{center}
	\includegraphics[width=0.4\textwidth]{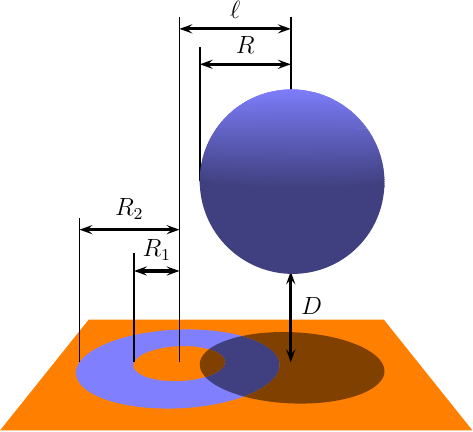}
	\caption{\label{fig:model}
	\ftitle{Sketch of a colloid above a substrate with a bull's-eye pattern.}
	The colloid and the substrate are immersed in a binary liquid mixture close to its critical demixing point. 
	We consider substrates and colloids with opposite boundary conditions denoted by orange and blue colors in this cartoon. 
	The bull's-eye pattern consists of a circular patch encircled by a ring with inner radius $R_1$ and outer radius $R_2$. 
	The boundary conditions inside the circular patch and outside the ring are the same (orange in the cartoon), but they are opposite to the boundary conditions inside the ring, where the chemical properties of the surface are the same as at the surface of the colloid (blue in the cartoon).
	The colloid is located at a surface-to-surface distance $D$ from the substrate and is shifted with respect to the center of the pattern by the distance $\ell$. The greyish circular patch to the right indicates the projection of the colloid onto the surface of the substrate. 
}
\end{center}
\end{figure}

We consider a colloidal particle above a patterned surface, as shown in \cref{fig:model}. The particle and the surface are immersed in a binary liquid mixture. In \cref{fig:model}, the preferences of the walls for different components of the liquid mixture are marked with blue and orange, respectively; for reasons of simplicity, we focus on the strong adsorption limit for the surfaces of both the substrate and the particle. 
We assume that the binary liquid mixture is close to its (lower) critical demixing point \cite{hertlein_direct_2008}: The composition is the critical one, and the temperature $T$ is close to its critical value $\Tc$. The radius of the colloidal particle is denoted by $R$ and its surface-to-surface distance to the patterned surface is $D$. Accordingly, the center of the colloidal particle is located at the distance $D+R$ above the surface. 

We consider a pattern consisting of two concentric circles with radii $R_1$ and $R_2$ (bull's-eye), which separate regions with the same and the opposite surface preferences as that of the colloid, aimed to provide a stable equilibrium trapping of the colloid in one or more directions.
The distance between the center of the orthogonal projection of the colloid onto the surface and the center of the pattern is denoted by $\ell$. Since the pattern exhibits rotational symmetry, there is no need to consider separately two coordinates of the projection of the center of the spherical particle.

Throughout the paper, we focus on the experimentally relevant case of temperatures below the lower critical demixing temperature $\Tc$, where the binary liquid mixture remains homogeneous.
These temperatures can be experimentally realized and maintained in a stationary state. Moreover, increasing the temperature towards criticality is generally easier to achieve compared to decreasing the temperature, as in the case of the upper critical point.

We use the temperature scaling variable $\Gamma= R/\xib\left(T\right)\cong R \; t^{\nu}/\xi_0^+$, where $t=\left(\Tc-T\right)/\Tc$ is the reduced temperature for the lower critical demixing point (note that $T<\Tc$ corresponds to $t > 0$ because the binary liquid mixture is homogeneous below $\Tc$), and $\xib\left(T\right)$ is the bulk correlation length, which diverges upon approaching the critical point with the non-universal amplitude $\xi_0^+$ and the universal critical bulk exponent $\nu$. 

In the scaling limit, \ie, $t\to 0$ with $\Gamma=R/\xib$, $\Delta=D/R$, $\Lambda=\ell/R$, $\rho_1=R_1/R$, and $\rho_2=R_2/R$ fixed, the potential $\CCP$ of the critical Casimir interaction between the colloid and the substrate is
\begin{equation}\label{scaling_law}
\CCP\left(D, \ell; R_1,R_2, R, T\right)=\kB \Tc\, \sfp\left(\Delta, \Lambda; \; \rho_1, \rho_2, \Gamma\right),
\end{equation}
where $\sfp$ is a universal scaling function, and $\kB$ denotes the Boltzmann constant. 
The position of the colloid is characterized by the coordinates $\left(D,\ell\right)$ or $\left(\Delta, \Lambda\right)$ with three fixed, dimensionless parameters $\left(\rho_1, \rho_2, \Gamma\right)$. In an equilibrium configuration, the coordinates become functions of these parameters.
We note that in the limit $\Delta = D/R \to 0$ the scaling function $\sfp$ diverges as $\Delta^{-1}$  \cite{Burkhardt1995}. 

From the potential, we calculate the force acting on the colloidal particle. Due to the symmetry of the system, the force has two independent components: $\fper=-\partial \CCP/\partial D$ acting in the direction perpendicular to the substrate, and $\flat=-\partial \CCP/\partial \ell$ acting in the direction parallel to the substrate, along the line perpendicular to the axis of symmetry of the pattern on the substrate, running through the centers of the orange and greyish patches in \cref{fig:model}.
In the scaling limit, the components of the force are given by the universal scaling functions
\begin{subequations}
\begin{align}
\fper\left(D, \ell; R_1,R_2, R, T\right)&=\frac{\kB \Tc}{R}\sffp\left(\Delta, \Lambda; \; \rho_1,\rho_2,\Gamma\right), \\
\flat\left(D, \ell; R_1,R_2, R, T\right)&=\frac{\kB \Tc}{R}\sffl\left(\Delta, \Lambda; \; \rho_1,\rho_2,\Gamma\right).
\end{align}
\end{subequations}
We note that the scaling functions are related according to $\sffp=-\partial \sfp/\partial \Delta$ and $\sffl=-\partial \sfp/\partial \Lambda$.
All three scaling functions ($\sfp$, $\fper$, $\flat$) are expected to be universal, \ie, they are identical for all systems belonging to the same universality class. 

\section{Methods}
\label{sec:methods}

There are no exact methods available to calculate the scaling functions $\sfp$, $\sffp$, and $\sffl$ systematically for the experimentally relevant universality class. Therefore, in order to study the levitation of the colloid, we resort to two approximate approaches, \viz, the Derjaguin approximation and MFT. The advantages of the Derjaguin approximation are that it is computationally fast and the results are free of finite-size corrections to scaling. The main disadvantage is that the accuracy of this approximation is hardly controllable. In general, one expects this method to give reliable results if a colloid is close to a substrate and 
far from any abrupt changes in the lateral boundary conditions \cite{FarahmandBafi2020}. Another inaccuracy stems from the scaling functions for the slab geometry $\varphi_\mathrm{s}$. Their error has been estimated to be about 20\% \cite{LabbeLaurent2016}. However, they agree quite well with experimental results \cite{hertlein_direct_2008,paladugu_nonadditivity_2016,magazzu_controlling_2019,wang_nanoalignment_2024-1}.

While MFT calculations lift the simplifications of the Derjaguin approximation (such as ignoring the free energy associated with the gradient of the order parameter in the directions parallel to the wall), they are computationally challenging, particularly in the close vicinity of the critical point, requiring large computational boxes and preventing us from performing calculations within a wide temperature window. In addition, MFT neglects fluctuations. We note that, in this respect, MFT yields the lowest order of a systematic expansion in terms of $\epsilon= 4-d$ of universal quantities such as critical exponents and scaling functions. Therefore, our mean-field analysis is an essential first step within a general scheme.

\subsection{Derjaguin approximation}

The Derjaguin approximation \cite{Derjaguin1934} estimates the interaction potential $\sfp$ of a spherical colloid above a flat surface by an integral:
\begin{equation}\label{DA:general}
    \pot=\int_S\dd x\, \dd y\, \mathfrak{u}\left(l\left(x,y\right)\right),
\end{equation}
where the integration is carried out over the circular disc $S$ of radius $R$, which is the orthogonal projection of the colloid onto the surface, $\dd x\,\dd y$ is the area element, and $l\left(x,y\right)$ is the spatially varying local separation between the substrate surface and the surface of the sphere (measured in the direction perpendicular to the planar surface). The function $\mathfrak{u}\left(l\right)$ is the energy of interaction per area calculated for the corresponding slit system with parallel homogeneous walls separated by a distance $l$. (The slit walls have the same properties as the local properties of the substrate surface and as the colloid surface at the point $\left(x,y\right)$.) 

In the case of critical Casimir interactions and in the scaling limit, within the slit system the free energy of interaction per area of the slit wall depends only on the separation $l$ of the walls and on the reduced temperature $t$:
\begin{equation}\label{DA:slab}
    \mathfrak{u}\left(l, t\right)=\frac{\kB \Tc}{l^2} \varphi_\mathrm{s}\left(\omega\right), 
\end{equation}
where $\omega = l/\xi \cong l \left|t\right|^\nu/\xi_0^+$ is a scaling variable, $\varphi_\mathrm{s}$ is a scaling function, and $\xi_0^+$ is a non-universal amplitude. For the Ising universality class, the bulk critical exponent $\nu$ is $\nu\approx0.63$ \cite{pelissetto_critical_2002} in $d=3$ (and $\nu=1/2$ within MFT, \ie, $d=4$). The index `$\mathrm{s}$' in \cref{DA:slab} denotes the so-called film universality class and, in our case, it is either `$\mathrm{sm}$' for ``same'' boundary conditions (\ie, both walls prefer the same component of the binary liquid mixture) or `$\mathrm{op}$' for ``opposite'' boundary conditions (\ie, the two walls prefer different components of the binary liquid mixture).

Using \cref{DA:general,DA:slab} in \cref{scaling_law}, we derive the expression for the scaling function for the free energy of interaction of the colloid with the patterned surface:
\begin{equation}\label{DA:formula}
    \sfp\left(\Delta, \Lambda; \rho_1, \rho_2, \Gamma\right)= \int_S \dd X\,\dd Y\,\left(\Delta+1-\sqrt{1-\left(X-\Lambda\right)^2-Y^2}\right)^{-2} \varphi_\mathrm{s}\left(\Omega\right),
\end{equation}
with $\Omega=\Gamma\left(\Delta+1-\sqrt{1-\left(X-\Lambda\right)^2-Y^2}\right)$, $S=\left\{\left(X,Y\right)\in \mathbb{R}^2\, \Big| \left(X-\Lambda\right)^2+Y^2\leqslant 1 \right\}$, and the film universality class `$\mathrm{s}$' is `$\mathrm{sm}$' for $\rho_1^2<X^2+Y^2<\rho_2^2$ and `$\mathrm{op}$' otherwise. In this way, $\rho_1$ and $\rho_2$ enter into the scaling function $\sfp$. 

We calculated the scaling function $\sfp$ within the Derjaguin approximation by carrying out the integration in \cref{DA:formula} numerically. To this end, we wrote a computer program in C++  using the GSL library \cite{Gough2009} for numerical integration. Since the exact and explicit formulae for the scaling functions for the slab geometry, $\varphi_\mathrm{sm}\left(\omega\right)$ and $\varphi_\mathrm{op}\left(\omega\right)$, are unknown, instead we have used their approximation based on Monte Carlo simulations for the slab system \cite{Gambassi2009,vasilyev_universal_2009,*vasilyev_erratum_2009,FarahmandBafi2020}. The scaling functions for the force, $\sffp$ and $\sffl$, can be obtained by numerically differentiating the interaction potentials. Alternatively, the function $\sffp$ can be obtained by using the Derjaguin approximation for the forces \cite{FarahmandBafi2020}.

\subsection{Mean field theory}
\label{sec:methods:mft}

We describe a critical fluid in spatial dimension $d$ by the standard dimensionless Landau--Ginsburg--Wilson (LGW) Hamiltonian:
\begin{align}
\label{eq:mft:H}
	\H[\varphi] = \int_V \dd^d r \left\{\frac{1}{2}(\nabla \varphi)^2 + \frac{\tau}{2} \varphi^2 + \frac{g}{4!}\varphi^4 \right\} ,
\end{align}
where $\dd^d r=\prod_{i=1}^{d} \dd x_i$ is the $d$-dimensional volume element, $V$ is the volume accessible to the fluid, and the coupling constant $g>0$ stabilizes $\H$ below the (upper) critical point $\Tc$. The order parameter $\varphi(\boldsymbol{r})$ is the difference between the local concentration of one of the components of the mixture and its concentration at the critical demixing point.
Within MFT one has $\tau = t/(\xi_0^+)^2$ where, as before, $t$ is the reduced temperature and $\xi_0^+$ the non-universal amplitude of the bulk correlation length $\xi\left(T\right)=\xi_0^{+}|t|^{-\nu}$ in the disordered phase. 
\label{ref:2:dim}We also assume that the system is translationally invariant in $d-3$ dimensions, which implies that the order parameter varies only in three spatial dimensions. The profile of the order parameter $\varphi\left(\boldsymbol{r}\right)$ is calculated by numerically minimizing the LGW Hamiltonian in \cref{eq:mft:H}. 

We consider the substrate and the colloid surface to belong to the so-called extraordinary (or normal) surface universality class, which corresponds to the limit of infinitely strong surface fields. (We recall that we also consider the strong adsorption limit in the calculations based on the Derjaguin approximation.)
This implies that the equilibrium concentration profile diverges upon approaching the surfaces (see, \eg, \myrefs{Binder1983, Diehl1986, Mohry2010}). In order to deal with this numerical challenge, we used a short-distance expansion (see \myrefs{Kondrat2007, Kondrat2009}) for calculating the value of the order parameter at a small distance $\delta$ from the surface (we took $\delta/R = 0.05$ in all calculations), and we kept these values fixed during the course of minimization. This approach has proven to be successful in several previous studies~\cite{Kondrat2009, Troendle2009, Troendle2010, mohry_critical_2014, labbe-laurent_alignment_2014, labbe-laurent_liquid_2017, Vasilyev2018, Kondrat2018, Vasilyev2020}. We used the three-dimensional finite element library F3DM~\cite{Kondrat2014} for minimizing numerically the LGW Hamiltonian (\cref{eq:mft:H}). In order to calculate the force acting between the colloid and the substrate, we enclosed the colloid by an ellipsoidal surface and computed the force by integrating the stress tensor over this surface \cite{Kondrat2009}.

\section{Results and discussion}

\subsection{General considerations}
\label{sec:res:general}

\begin{figure}
\begin{center}
	\includegraphics[width=\textwidth]{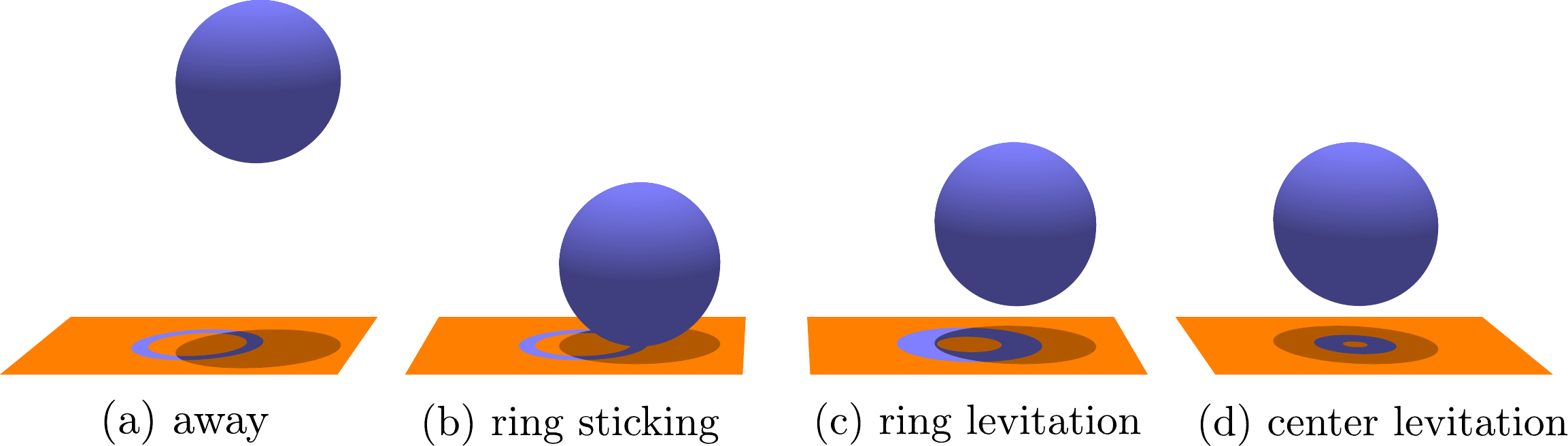}
    \caption{\label{fig:schematics}
    \ftitle{Four (meta)stable configurations of a colloid near a substrate with a bull's-eye pattern.} This schematic drawing (compare \cref{fig:model}) shows that a colloid can 
	\fsub{a} stay macroscopically far away from the substrate, 
	\fsub{b} stick to the ring of the bull's-eye pattern,
	\fsub{c} levitate above the ring, and
	\fsub{d} levitate above the substrate at the center of the bull's-eye pattern. 
 	We note that the opposite boundary conditions prevent the colloid from sticking to the center of the bull's-eye pattern.
        In configuration (a), the colloid is not confined in the lateral directions, while in configurations (b) and (c), it is free to diffuse along the ring.
}
\end{center}
\end{figure}

Typically, if two surfaces confining a critical fluid have similar boundary conditions, they attract each other, whereas two surfaces with opposing boundary conditions repel each other.
Therefore, varying the bull's-eye radii $R_1$ and $R_2$, as defined in \cref{fig:model}, may allow one to control the sign and the strength of colloid--wall interactions.
This results in distinct stable or metastable positions of a colloid above the surface. Before discussing how these positions depend on $R_i$ ($i=1,2$) and temperature, we present general considerations concerning the behavior one can expect as a function of the bull's-eye parameters. 
There are four possible types of configurations (we note that some of them may coexist under specific thermodynamic conditions, as will be discussed below). 
These configurations are illustrated in \cref{fig:schematics} and described as follows:
\renewcommand{\labelenumi}{(\alph{enumi})}
\begin{enumerate}
\item 
\label{ref:2:q5}For sufficiently narrow rings, we expect the wall to repeal the colloid for all lateral shifts at distances larger than the distance at which, for the colloid above the ring, the attraction from the ring counter-balances the repulsion from the rest of the surface. This implies that $D\to \infty$ corresponds to a local minimum of the colloid--wall potential. 

In the absence of other interactions, this local minimum does not correspond to the global one.
The corresponding configuration is shown schematically in \cref{fig:schematics}(a). In this configuration, the colloid is not confined in the lateral directions.

\item The colloid may stick to the ring part of the surface pattern (\cref{fig:schematics}(b)). In the limit of strong adsorption, as considered here, this configuration represents the global minimum with an unlimited depth of the interaction potential (cut off by a hard-core repulsion), independently of other parameters. We note that electrostatic interactions, which are almost always present in such systems, may experimentally prevent sticking to the ring or lead to ring levitation (see \cref{fig:schematics}(c)).

\item If the attraction to the ring balances the repulsion from the rest of the wall, the colloid may levitate above the ring (\cref{fig:schematics}(c)). We note that the colloid can move freely along the ring. For large radii ($R_i \gg R$), this configuration is similar to levitation above a stripe, studied by \citeauthor{Troendle2009} \cite{Troendle2009, Troendle2010, Troendle2011}.

\item Similarly, if the combined repulsion from the bull's-eye center and the rest of the surface (\ie, from the orange region on the surface) balances the attraction to the ring, the colloid can levitate above the center of the pattern (\cref{fig:schematics}(d)). We are primarily interested in this `point-like' levitation of the colloid.
\end{enumerate}

While configuration \confRing (\cref{fig:schematics}(b)) is always present and corresponds to the global minimum, the configurations \confInfty, \confRlev, and \confPlev (\namecrefs{fig:schematics}~\ref{fig:schematics} (a),(c), and (d)) may or may not exist, depending on the bull's-eye radii and temperature.  
We note that the characterization of configuration \confInfty requires the potential to be repulsive at large separations from the substrate. However, in practice, far away from the substrate the critical Casimir interaction is negligible compared to $\kB T$ so that the particle can stay out there regardless of the sign of the potential.

\subsection{Configuration diagram within the Derjaguin approximation}
\label{sec:res:conf_diagram}

\begin{figure}
\begin{center}
	\includegraphics[width=1.05\textwidth]{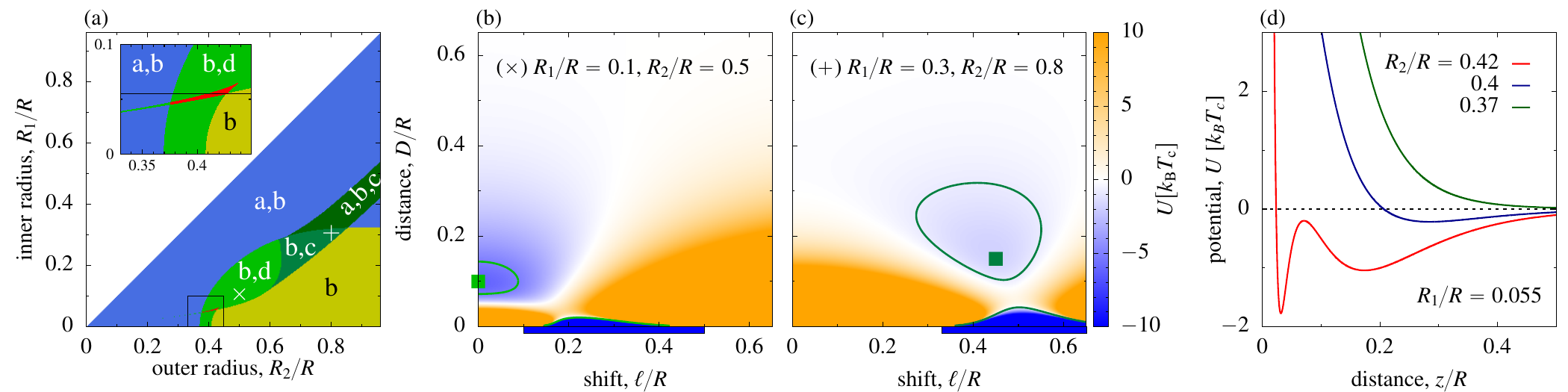}
    \caption{\label{fig:conf_diagram}
    \ftitle{Configuration diagram within the Derjaguin approximation.} 
	\fsub{a} Configuration space for a spherical colloid above a surface with a bull's-eye pattern, calculated within the Derjaguin approximation. The letters `a', `b', `c', and `d' in the plot correspond to the types of (meta)stable configurations presented in \cref{fig:schematics}.
	The green colors of different shades denote the regions of particle levitation (configurations `c' and `d').
	The symbols $\times$ and $+$ indicate the values of the radii $R_1/R$ and $R_2/R$ used for the panels (b) and (c), respectively.
	The inset details the region indicated by the rectangular box.
	The red color denotes the region with two coexisting minima (see panel (d)), and the horizontal black line indicates the value of $R_1/R$ used in panel (d).
	\fsub{b,c} Heatmaps of the interaction potential between the colloid and the surface in the plane spanned by the reduced surface-to-surface distance $D/R$ and the reduced shift $\ell/R$ relative to center of the bull's-eye pattern.
	The filled squares show the locations of the minima corresponding to point levitation ($\ell=0$, $D>0$) and ring levitation ($\ell>0$, $D>0$), respectively. 
	The greenish contour lines show the values of $\ell/R$ and $D/R$ at which the potential is $k_BT_c$ higher than at the corresponding minima.
	The blue rectangular bars beneath the horizontal axes mark the location of the ring with the same boundary condition as the colloid (see \cref{fig:model,fig:schematics}).
	\fsub{d} Interaction potential between the colloid and the surface as a function of the distance from the surface for $R_1/R = 0.055$ and for three values of $R_2/R$.
	The temperature scaling variable is $\Gamma= R/\xi=10$ in all panels.}
\end{center}
\end{figure}

We used the Derjaguin approximation in order to investigate under which conditions these four configurations are (meta)stable. 
For simplicity, and because we are primarily interested in center levitation, which allows one to localize colloids both in the vertical and lateral directions (\ie, at $\ell= 0$, so far not investigated in the literature), we studied the occurrence of local minima far away from the substrate (configuration \confInfty in \cref{fig:schematics}) only for $\ell=0$.

\Cref{fig:conf_diagram}(a) shows our results in terms of a configuration diagram in the plane spanned by the two bull's-eye radii $R_1$ and $R_2 > R_1$, and for the temperature scaling variable $\Gamma=R/\xib=10$. 
The various regions in this diagram are denoted by configurations (a, b, c, and d, see \cref{fig:schematics}) which are (meta)stable in the corresponding regions.
We first consider the blue region below the line $R_1=R_2$ in \cref{fig:conf_diagram}(a).
In this case, for $R_1 \lesssim R_2$, the ring is so thin that it cannot attract a particle from large distances in spite of equal boundary conditions. 
Therefore, only two configurations are possible: The particle is far away from the surface or the particle is stuck to the ring (configurations \confInfty and \confRing in \cref{fig:schematics}). 

If we take $R_1 \gtrsim 0.325R$ and increase $R_2$ (\ie, the outer radius), we enter the dark green region in \cref{fig:conf_diagram}(a) (denoted as ``a,b,c''), which is characterized by a minimum in the interaction potential at $D = \Dpot < \infty$ and nonzero $\lpot$. This amounts to ring levitation (configuration \confRlev in \cref{fig:schematics}). At the left boundary of this region, this minimum is shallow and located far away from the substrate. With increasing $R_2$, $\Dpot$ decreases and the minimum gets deeper. Upon increasing $R_2$ further, the barrier between ring levitation and sticking to the substrate decreases and eventually vanishes; hence, there is no levitation configuration in the blue region on the right-hand side of the dark green area in \cref{fig:conf_diagram}(a), similarly as in the blue region on the left-hand side of this area (\ie, for $R_2 \gtrsim R_1$).

We now choose the values of $R_1$ and $R_2$ such that they correspond to ring levitation, and we increase both $R_1$ and $R_2$ simultaneously, \ie, we move along the dark green stripe denoted as ``a,b,c'' in \cref{fig:conf_diagram}(a). In doing so, we asymptotically approach the case of levitation above a straight stripe studied by \citeauthor{Troendle2011} \cite{Troendle2009, Troendle2010, Troendle2011}. 
If, instead, we decrease $R_1$ and $R_2$ (around $R_1\approx 0.325 R$), the circle in the middle of the pattern becomes too small to induce repulsion at large distances and, for sufficiently large $R_2$, the minimum far away (configuration \confInfty in \cref{fig:schematics}) vanishes.
We thus enter the medium-dark green region denoted as ``b,c'' in \cref{fig:conf_diagram}(a).
Upon reducing $R_1$ and $R_2$ further, the shift $\lpot$ of the potential minimum decreases, and the ring levitation transforms into a point levitation (see configuration \confPlev in \cref{fig:schematics} and the light green region denoted as ``b,d'' in \cref{fig:conf_diagram}(a)) in a manner which resembles a continuous morphological transition. Examples of the potential energy surface for point ($\lpot=0$, $\Dpot>0$) and ring ($\lpot>0$, $\Dpot>0$) levitations are shown in \namecrefs{fig:conf_diagram}~\ref{fig:conf_diagram}(b) and (c). 

There are three scenarios for which point levitation can disappear upon varying the bull's-eye radii:
\begin{enumerate}
	\item[(i)] Keeping $R_1\gtrsim 0.06R$ fixed and increasing $R_2$ leads to spontaneous symmetry breaking which changes point levitation to ring levitation, as described above.
	\item[(ii)] By keeping $R_1\gtrsim 0.06R$ fixed and decreasing $R_2$, the minimum at $\ell=0$ becomes less deep, moves away from the substrate, and eventually disappears at far distances $D$.
	\item[(iii)] With reducing $R_1$ for $R_2\gtrsim0.43R$, the free energy barrier between point levitation and sticking to the substrate decreases,  and the minimum eventually turns into a saddle point. 
\end{enumerate}

If $R_1$ is reduced for $R_2\lesssim 0.43R$, we enter the region in which the potential for $\ell=0$ has two local minima at different colloid--wall separations, which correspond to two distinct point levitation configurations (see the narrow red region in the inset of \cref{fig:conf_diagram}(a)).

In the middle of this region is a line along which these two configurations co-exist. (We could not precisely locate it because this region is very narrow.) 
This line corresponds to a first-order morphological transition between the two configurations and ends at a critical point, at which the difference between them vanishes (\ie, the two distances from the wall, at which we observe the two minima, merge); we estimate that this morphological critical point occurs at $R_1^\ast\approx 0.06R$ and $R_2^\ast\approx0.43R$.
Thus, the red region in \cref{fig:conf_diagram}(a) is a spinodal region like any spinodal close to a critical point, as in the case of liquid--gas phase transitions or phase segregations in binary-liquid mixtures \cite{binder_theory_1987}.
Consistently, the shape of the potential around the morphological critical point indeed resembles the one of the Landau free energy \cite{landau_theory_1937, yukhnovskii_phase_1987}.

The two minima, present in the red (spinodal) region in \cref{fig:conf_diagram}(a), persist outside of this region. 
One minimum extends to smaller values of $R_2$ for a very narrow range of $R_1$. It is beyond the resolution of \cref{fig:conf_diagram}(a), but part of it is visible in its inset---see the narrow green stripe in the blue (``a,b'') region on the left side of the inset. This stripe becomes narrower with decreasing $R_1$ and $R_2$ and likely approaches the point $R_1=R_2=0$ (the extreme narrowness of this region prevented us from investigating its behavior in more detail).
The second minimum extends to smaller values of $R_1$ (the light green region below the red stripe, see the inset in \cref{fig:conf_diagram}(a)).
Both minima exhibit very small potential barriers which separate them from other configurations (\viz, the configuration of the particle sticking to the substrate and of the particle being far away from the substrate). 
Therefore, we refrain from discussing this case further.

\subsection{Point levitation obtained from mean-field calculations}
\label{sec:res:mft}

\begin{figure}
\begin{center}
	\includegraphics[width=0.85\textwidth]{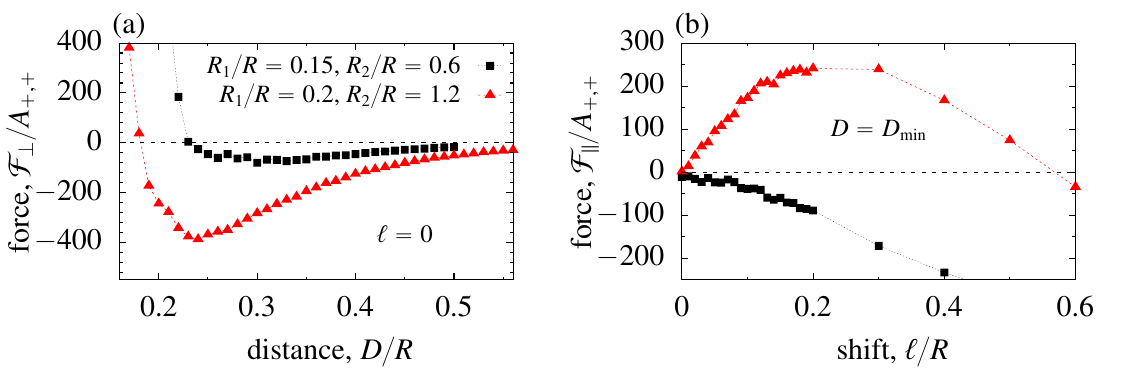}
	\caption{\label{fig:mft}
	\ftitle{Mean-field results for colloid levitation.}
	\fsub{a} Scaling function $\sffp$ of the force acting on the colloid in the direction perpendicular to the substrate \versus surface-to-surface distance $D/R$ for zero shift ($\ell=0$). 
	\fsub{b} Scaling function $\sffl$ of the force acting on the colloid in the plane $D=\Dmin$ parallel to the substrate \versus the shift $\ell/R$; the case $D=\Dmin$ corresponds to zero force at zero shift. The forces are expressed in terms of the critical amplitude $A_{+,+}$ of the plate--plate force scaling function for equal boundary conditions.
	The radii $R_1$ and $R_2$ of the inner and outer rings, respectively, are indicated in the plot (panel (a)). 
	  Positive values of $\sffp$ mean that the colloid is repelled from the wall, and $\sffl>0$ denotes repulsion from the center of the pattern. 
	The red lines and triangles belong to a saddle point, while the black lines and squares provide an example of a local minimum, corresponding to a colloid levitating above the bull's-eye center.
	The temperature scaling variable is $\Gamma=R/\xib=10$ for both plots.
}
\end{center}
\end{figure}

Since numerical mean-field calculations are computationally expensive and the stripe levitation---a limiting case of the ring levitation---has already been studied within MFT by \citeauthor{Troendle2010} \cite{Troendle2009, Troendle2010}, here we only explore the possibility of point-like levitation above the bull's-eye center. In \cref{fig:mft}, we present mean-field results for two examples of distinct inner and outer radii. 
Instead of plotting the interaction potentials, as in \namecrefs{fig:conf_diagram}~\ref{fig:conf_diagram}(b)--(c), we show the scaling functions $\sffp$ and $\sffl$ for the forces acting on the colloid perpendicularly to the surface (\cref{fig:mft}(a)) and in the lateral direction away from the bull's-eye center (\cref{fig:mft}(b)), respectively. 

We first computed the scaling function $\sffp$ of the perpendicular force for zero shift ($\ell=0$).
In both cases presented in \cref{fig:mft}(a), one has $\sffp>0$ at small distances $D$, \ie, the force is repulsive due to the repulsion exerted by the bull's-eye center. 
However, since the rings are sufficiently wide, they induce an attraction towards the substrate at large distances. 
Thus, $\sffp$ vanishes at a certain intermediate distance $D=\Dmin$, which corresponds to the minimum of the potential, indicating the possibility of point-like levitation.

For a colloid to levitate above the bull's-eye center, its configuration must be stable also with respect to lateral shifts $\ell$. 
\Cref{fig:mft}(b) shows the $\ell$-dependence of the scaling function  $\sffl$ of the lateral force for the two patterns presented in \cref{fig:mft}(a) at the surface-to-surface distances $\Dmin$, which correspond to $\sffp=0$ at $\ell=0$. 
For the wider circles (red symbols in \cref{fig:mft}(b)), the lateral force is positive for small non-zero shifts. 
This implies that the colloid is repelled from the bull's-eye center,
\ie,
this case corresponds to a saddle point in the interaction potential at $D=\Dmin$ and $\ell=0$. For the narrower circles (black symbols in \cref{fig:mft}), one has $\sffl < 0$ for small non-zero $\ell$, which means that the position of the colloid is stabilized at zero shift. In this case, the colloid levitates above the center of the bull's-eye pattern (\cref{fig:schematics}(d)). Accordingly, the condition for point levitation is $\sffp=0$ for $\ell=0$ and $\sffl>0$ for small non-zero $\ell$. 
We used this condition in order to determine the levitation diagram within MFT (see the next section).

\subsection{Levitation diagram}
\label{sec:res:lev_diagram}

\begin{figure}
\begin{center}
	\includegraphics[width=\textwidth]{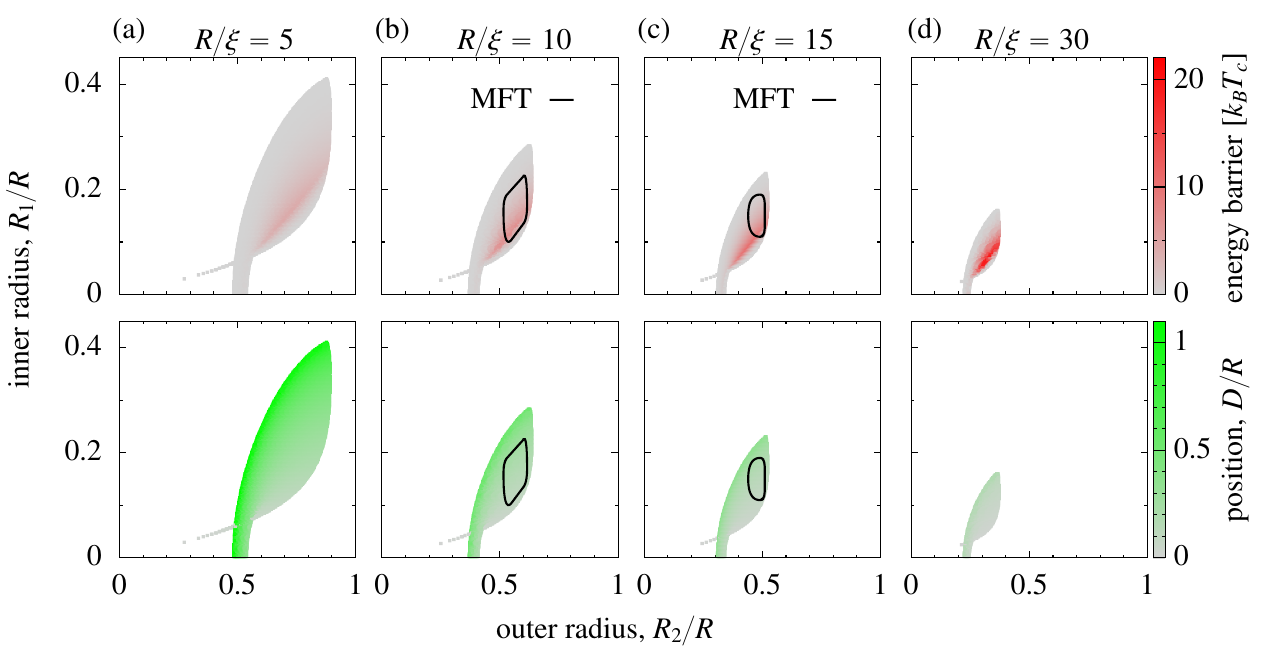}
	\caption{\label{fig:diagram}
    	\ftitle{Levitation diagram.} 
	Heatmap plots of the free energy barriers for point levitation (top row, $\ell=0$) and surface-to-surface distance between  the substrate and the levitating colloid (bottom row) for four values of the temperature scaling variable $R/\xib\approx R\left| t\right|^{\nu}/\xi_0^+$, where $t=\left(\Tc-T\right)/\Tc>0$ is the reduced temperature, as indicated in the plot. 
	The data have been calculated within the Derjaguin approximation. 
	The solid lines in the panels (b) and (c) enclose the regions of point levitation as obtained within MFT.
	The shape of the plots of the free energy barriers and of the colloid positions correspond to the shape of the b,d region in \cref{fig:conf_diagram}(a) where the potential has at least one minimum. 
\label{ref:2:fig5}
	In all panels, we keep the same scale for the outer radius $R_1$ and for the inner radius $R_2$ in order to emphasize the shrinking region of point levitation as the temperature scaling variable $R/\xi$ increases, \ie, as the system is taken away from criticality.
}
\end{center}
\end{figure}

For levitation to be possible, the corresponding configuration must be stable against the thermal fluctuations present in the system. In order to address this issue, at least partially, we used the Derjaguin approximation to calculate the potential barriers separating the minimum corresponding to the point levitation from the minima of other configurations (\cref{fig:schematics,fig:conf_diagram}):

\begin{itemize}
    \item 
    
	In order to calculate the free energy barrier between the point levitation and a colloid far away (\ie, between the configurations \confPlev and \confInfty in \cref{fig:schematics}), it turned out to be sufficient to consider the potential for $\ell=0$. Therefore, we determined the maximal value of the potential as a function of $D$ for $D\geqslant \Dmin$, and computed the free energy barrier as the difference between this maximal value of the potential and the value of the potential corresponding to the point levitation (\ie, at $D=\Dmin$).

	\item 
	In order to estimate the free energy barrier between point levitation and sticking to the substrate (\ie, between the configurations \confPlev and \confRing in \cref{fig:schematics}), we constructed a family of parabolic curves in the $(D, \ell)$ plane, connecting the point levitation configuration ($D=\Dmin$, $\ell=0$) with the point which represents sticking to the ring (\ie, $D=0$, $\ell=\left(R_1+R_2\right)/2$), parameterized by a certain parameter $\alpha$. For each value of $\alpha$, we obtained the maximal value of the potential along the curve and then minimized these maximal values with respect to $\alpha$. This procedure allowed us to locate a saddle point separating the two minima corresponding to point levitation and sticking to the ring. The free energy barrier equals the difference between the value of the potential at the saddle point and the point levitation minimum.

    \item 
	In all cases considered, we have not found system parameters which simultaneously provide point and ring levitation (configurations \confPlev and \confRlev in \cref{fig:schematics}). Therefore, there was no need to calculate a barrier between these two configurations.
\end{itemize}

The free energy barriers plotted in \cref{fig:diagram} as heatmaps correspond to the smallest free energy barriers as determined above. In the case of having two minima, which both correspond to point levitation (red region in \cref{fig:conf_diagram}), we present the data for the minimum with the highest barriers. 
On the bottom panels, we also show the corresponding surface-to-surface distances between the substrate and the levitating colloid.
These `levitation diagrams’, as we call them, are drawn in the plane of the bull's-eye radii $R_1/R$ and $R_2/R$. The shapes of the heatmaps, characterized by zero free energy barriers, correspond to the region of point levitation shown in \cref{fig:conf_diagram}(a). This region decreases as one moves away from critical point, \ie, upon increasing $\Gamma=R/\xib$.
However, simultaneously, the maximum observed height of the potential barrier increases (reddish areas in \cref{fig:diagram}), \ie, with increasing $R/\xib$ the particle becomes confined more strongly.
This increase in the free energy barrier is accompanied by the decrease of the levitation distance $\Dmin/R$, which occurs because the correlation length decreases with increasing $\Gamma=R/\xib$.
For small $R/\xib$ (\ie, closer to $\Tc$), the repulsion from the bull's-eye center is much stronger than the attraction to the ring, which pushes the colloid away from the center, rendering the point levitation unstable at small distances.
Increasing $R/\xib$ allows one to counterbalance the repulsion while simultaneously the difference between the strengths of the repulsion and of the attraction increases. Accordingly, the confining potential is strengthened, which facilitates point levitation at small distances.

By applying mean-field calculations we validated the levitation diagrams (top panels in \cref{fig:diagram}) obtained by using the Derjaguin approximation.
The mean-field results are shown as solid lines in \namecrefs{fig:diagram}~\ref{fig:diagram}(b) and (c) for two temperatures ($\Gamma=10$ and $\Gamma=15$). For smaller ratios $R/\xib$, the numerical MFT calculations become challenging due to the large correlation length and hence due to the associated large size of the computational boxes. At large $R/\xib$, the point levitation region is small, making it challenging to locate it. 
The boundary of the region of point levitation is determined by (i) the vanishing of the force in the direction perpendicular to the substrate and (ii) by a negative (\ie, attractive) lateral force pointing into the direction of the bull's-eye center (see \cref{sec:res:mft}).
The regions determined this way are smaller than computed within the Derjaguin approximation.
However, the MFT results confirm the occurrence of point levitation and show that the region of point levitation decreases as $R/\xib$ increases, in line with the behavior of the Derjaguin approximation.

\subsection{Two possible applications of critical Casimir levitation}
\label{sec:res:appl}

We now discuss two potential applications of point levitation: (i) sorting colloids by size and (ii) measuring temperature near criticality.

\subsubsection{Sorting colloids by size}

\Cref{fig:diagram} illustrates that the region of point levitation depends sensitively on the value of the temperature scaling variable $\Gamma = R/\xi$. Since $\Gamma$ can be changed by altering the particle size or the correlation length (via temperature $T$), this figure suggests that, at a given $T$, only particles of certain sizes can be trapped above the bull's-eye pattern. This selective trapping can be utilized for sorting particles by their size.

In order to demonstrate it, we computed a levitation diagram within the Derjaguin approximation similar to that in \cref{fig:diagram}. Now, however, we set the outer radius $R_2$ of the bull's-eye pattern as a length scale and chose  $R_2/\xi$ as the temperature scaling variable. 
We then plot a diagram in the plane spanned by $R_1/R_2$ and $R/R_2$ for a fixed value of $R_2/\xi$.
In other words, we fix the temperature and the pattern size through $R_2/\xi$ and seek pattern configurations (\ie, $R_1/R_2$), which provide trapping for a particle of size $R/R_2$. 

\begin{figure}
\begin{center}
	\includegraphics[width=0.9\textwidth]{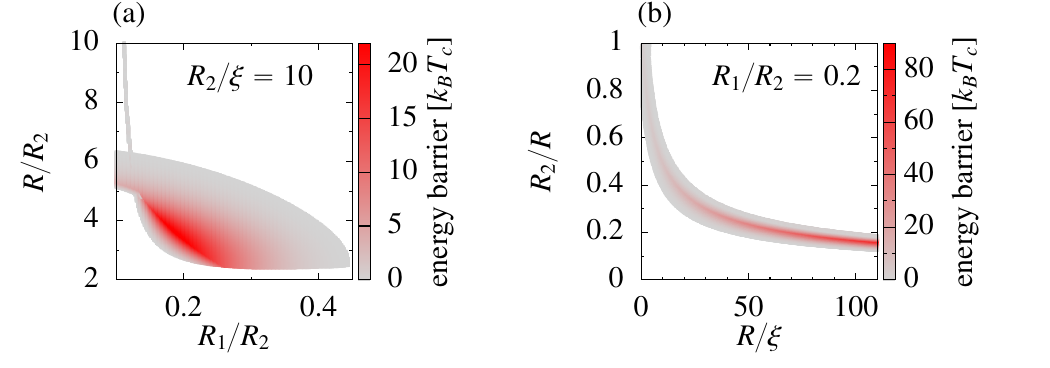}
	\caption{\label{fig:appl}
    	\ftitle{Applications of critical Casimir levitation.} 
	\fsub{a} Sorting particles by size: The heatmap shows the free energy barriers for point levitation with the temperature scaling variable $R_2/\xib\approx R_2\left| t\right|^{\nu}/\xi_0^+ = 10$, where $t=\left(\Tc-T\right)/\Tc>0$ is the reduced temperature.
	The heatmap is plotted in the plane of the ratio $R_1/R_2$ of the inner and outer radii of the bull's-eye pattern and of the colloid radius $R$ expressed in terms of the outer radius $R_2$.
	For a given ratio $R_1/R_2$, only colloids of limited sizes can levitate above the bull's-eye pattern.
	\fsub{b} Critical Casimir thermometer: The heatmap shows the free energy barriers for point levitation in the plane spanned by the temperature scaling variable $R/\xib$ and the pattern size $R_2/R$ for a fixed ratio $R_1/R_2$ of the bull's-eye radii.
	The thermodynamic distance to criticality can be measured by mapping the correlation length $\xib$ onto the range of bull's-eye sizes $R_2$ providing point levitation for a test particle of radius $R$.
	The data in both panels have been calculated within the Derjaguin approximation. 
}
\end{center}
\end{figure}

The results of our calculations are plotted in \cref{fig:appl}(a). This figure demonstrates that, for a given ratio $R_1/R_2$, only particles of specific sizes can levitate above the center of the bull's-eye pattern. By judiciously choosing the values of $\xi$, $R_2$, and $R_1$, the range of particle sizes can be adjusted to desired values, thus providing a means to sort particles by size.
As an example, referring to \cref{fig:appl}(a), we choose $R_2=\SI{1}{\micro\meter}$ and $\xib=\SI{100}{\nano\meter}$ (so that $R_2/\xib=10$) and take $R_1/R_2=0.05$. This choice leads to the trapping of colloids of sizes from ca.~\SIrange{5}{6}{\micro\meter}.

\subsubsection{Critical Casimir thermometer}

The temperature sensitivity of particle trapping can also be utilized to measure temperatures close to criticality. To demonstrate this possibility, we computed the levitation diagram in the plane ($R/R_2, R/\xi$) for a fixed ratio of the bull's-eye radii $R_2/R_1$. The results, plotted in \cref{fig:appl}(b), show that only patterns of specific sizes provide point levitation for colloids of size $R$ at a given temperature (\ie, for a fixed $R/\xi$). 
One can now create a surface pattern containing many widely separated bull's-eyes structures of various sizes (\ie, of different $R_2$) but with the same ratio $R_1/R_2$. 
This arrangement can be used to measure the temperature by mapping the correlation lengths $\xib$ to the range of bull's-eye patterns (\ie, to $R_2$) which allow point levitation for particles of radius $R$. 
Such a ``critical Casimir thermometer'' would be useful if direct measurements of the thermodynamic distance to criticality (\ie, the correlation length) are challenging.
Notably, the pattern sizes ($R_2$) providing point levitation exhibit a steep increase as $R/\xi$ decreases (\cref{fig:appl}(b)). This indicates the potential use of the critical Casimir thermometer near criticality (\ie, for $R/\xi \to 0$), at which conventional temperature measurements often lack accuracy.

As an example, we consider a test particle of radius $R=\SI{1}{\micro\meter}$. At a temperature corresponding to a correlation length of $\xib=\SI{10}{\nano\meter}$ ($R/\xib=100$), the test particles will be trapped above bull's-eye patterns with sizes ranging from ca.~\SIrange{125}{200}{\nano\meter}.
For $\xib=\SI{100}{\nano\meter}$, \ie, closer to the critical point ($R/\xib=10$), the same test particles will be trapped above patterns with sizes ranging from ca.~\SIrange{380}{600}{\nano\meter}. 

\section{Conclusions}
\label{sec:concl}

We have considered a bull's-eye chemical pattern on a planar substrate, which is composed of circular regions with antagonistic surface properties (\eg, opposite preferences for the two components of a binary liquid mixture; see \cref{fig:model}). 
We studied how to use critical Casimir forces to make colloidal particles levitate above this pattern. 
To this end, we employed the Derjaguin approximation and validated some results via full-fledged, numerical mean-field calculations.

We identified four configurations of a colloid above such a patterned surface, \viz, \confInfty if the colloid is far away from the surface, \confRing if it sticks to the bull's-eye ring, \confRlev if it levitates above the ring, and \confPlev if it levitates above the bull's-eye center, which we call point levitation (\cref{fig:schematics}). Within the Derjaguin approximation, we calculated a morphological phase diagram in the plane spanned by the two bull's-eye radii. It reveals a rich behavior with two or three coexisting configurations (\cref{fig:conf_diagram}). 
Within our model, sticking to the ring corresponds to the global free energy minimum, independently of the model parameters, while the other configurations are at most metastable.
We note, however, that electrostatic interactions, which virtually always are present in the corresponding experimental systems (see, \eg, \myrefs{trondle_trapping_2011,wang_nanoalignment_2024-1}), may impede sticking to the ring.
(The same boundary conditions would also mean the same surface charges,
which in turn would lead to repulsion.)

For ring levitation, colloidal particles stay some distance above the surface but preserve their ability for lateral diffusion along the ring---akin to levitation above surfaces with periodic stripe patterns, previously explored theoretically \cite{Troendle2009, Troendle2010, Troendle2011} and experimentally \cite{Troendle2011}.
We therefore focused on point levitation, which provides the confinement of colloids in perpendicular and lateral directions relative to the surface.
We found a sensitive dependence of the confining potential on temperature.
The domain of stability of point levitation shrinks as temperature increases (\ie, as the system is taken away from the bulk critical point). 
Surprisingly, however, the confining potential for levitation becomes deeper upon taking the system away from criticality, while the distance between the substrate and the levitating colloid decreases (\cref{fig:diagram}). This sensitive dependence of the confining potential on temperature might be utilized to sort colloids of different sizes or to construct a ``critical Casimir thermometer'' for  measuring the thermodynamic distance to criticality (\ie, the correlation length). 

Our theoretical predictions are within the range of current experimental possibilities \cite{wang_nanoalignment_2024-1}.
The substrates can be realized with relatively standard microfabrication techniques such as e-beam lithography, while the particle motion can be measured by microscopy techniques, such as dark-field or holographic microscopy.
Both point and ring levitation, as described in the present study, can potentially be achieved by using techniques similar to those employed in \myref{wang_nanoalignment_2024-1}. In fact, the manipulation of colloidal particles above patterned surfaces shares conceptual similarity with the control of microdisk alignment and positioning discussed in \myref{wang_nanoalignment_2024-1}. By adapting the  nanopatterning strategies and the temperature-dependent control developed in that study, one could enhance the capabilities of the bull's-eye patterns to control the vertical and lateral positioning of colloidal particles in an experimental setting.
 
\section*{Data Availability Statement}

The data that support the findings of this study are available from the corresponding author upon reasonable request. 

\begin{acknowledgements}

	We acknowledge scientific discussions with Dr.~O.~A.~Vasilyev regarding this project during his stay at MPI-IS in Stuttgart. 
	NFB  was  supported by  the Polish National Science Center (Opus Grant  No.~2022/45/B/ST3/00936).
\end{acknowledgements}

\bibliography{refs,bull_eye}

\end{document}

%% file: defines.tex
\def\papertitle{Critical  Casimir levitation of colloids above a bull's-eye pattern}

\usepackage{graphicx} 
\usepackage{amsmath}
\usepackage{amsfonts}
\usepackage{amssymb}
\usepackage{xcolor}

\usepackage[draft,inline,nomargin]{fixme}
\usepackage{grffile}

\usepackage{hyperref}
\usepackage{cleveref}
\crefname{figure}{Fig.}{Figs.}
\Crefname{figure}{Figure}{Figures}
\crefname{equation}{Eq.}{Eqs.}
\crefname{section}{Sec.}{Sec.}

\fxsetup{theme=color,mode=multiuser}
\FXRegisterAuthor{sk}{ask}{\color{orange}SK}
\FXRegisterAuthor{pn}{apn}{\color{cyan}PN}
\FXRegisterAuthor{nf}{anf}{\color{magenta}NFB}

\usepackage{siunitx}
\DeclareSIUnit{\calorie}{cal}
\DeclareSIUnit{\Calorie}{\kilo\calorie}
\usepackage{xspace}
\newcommand{\latin}[1]{{\it #1}}
\newcommand{\ie}{\latin{i.e.}\@\xspace}
\newcommand{\eg}{\latin{e.g.}\@\xspace}

\newcommand{\viz}{\latin{viz.}\@\xspace}

\newcommand{\versus}{\latin{versus}\@\xspace}

\newcommand{\ftitle}[1]{{\bf #1}}
\newcommand{\fsub}[1]{({\bf #1})}

\newcommand{\myref}[1]{Ref.~\cite{#1}}
\newcommand{\myrefs}[1]{Refs.~\cite{#1}}

\newcommand{\Tc}{T_\mathrm{c}}
\newcommand{\kB}{k_\mathrm{B}}
\newcommand{\xib}{\xi}
\newcommand{\pot}{\mathfrak{U}}
\newcommand{\fper}{\mathfrak{F}_\perp^\mathrm{c}}
\renewcommand{\flat}{\mathfrak{F}_\parallel^\mathrm{c}}
\newcommand{\CCP}{\pot^\mathrm{c}}
\newcommand{\dd}{\mathrm{d}}
\newcommand{\sign}{\operatorname{sign}}
\newcommand{\sfp}{\mathcal{U}}
\newcommand{\sffl}{\mathcal{F}_\parallel}
\newcommand{\sffp}{\mathcal{F}_\perp}

\renewcommand{\H}{\mathcal{H}} 

\newcommand{\Dmin}{D_\mathrm{min}}
\newcommand{\Dpot}{D_\mathrm{min}}
\newcommand{\lpot}{\ell_\mathrm{min}}

\def\confPlev{(d)\@\xspace}
\def\confRlev{(c)\@\xspace}
\def\confRing{(b)\@\xspace}
\def\confInfty{(a)\@\xspace}